# Modulating trap properties by $Cr^{3+}$-doping in $Zn_2SiO_4$: $Mn^{2+}$ nano phosphor for optical information storage

Xin Yi[1], Hui Liu[1], Yihuan Wang[1], Junjie Chen[1], Zhanglin Chen[3], Yuzhen Wang[4], Xuanyi Yuan[1,*], Kaiming Zhu[2,**]

1. *Beijing Key Laboratory of Optoelectronic Functional Materials & Micro-nano Devices, Department of Physics, Renmin University of China, Beijing 100872, P.R. China*
2. *College of Science, Hebei Agricultural University, Baoding 071001, P.R. China*
3. *State Key Laboratory of New Ceramics & Fine Processing, School of Materials Science and Engineering, Tsinghua University, Beijing 100084, P.R. China*
4. *State Key Laboratory of Luminescent Materials and Devices, Guangdong Provincial Key Laboratory of Fiber Laser Materials and Applied Techniques, Guangdong Engineering Technology Research and Development Centre of Special Optical Fiber Materials and Devices, School of Physics and Optoelectronics, South China University of Technology, Guangzhou 510641, P.R. China*

**Corresponding authors:**

Xuanyi Yuan: E-mail addresses: yuanxuanyi@ruc.edu.cn; postal address: Renmin University of China, No. 59 Zhongguancun Street, Haidian District, Beijing 100872 P.R. China.

Kaiming Zhu: E-mail address: zhukaiming312@163.com; postal address: Hebei Agricultural University, No. 289 Lingyu Temple Street, Lianchi District, Baoding 071001, P.R. China.

**Abstract:**

Photo-stimulated luminescent materials are one of the most attractive alternatives for next-generation optical information storage technologies. However, there are still some challenges in regulating appropriate energy levels in luminescent materials for optical information storage. Herein, a green emission nanophosphor $Zn_2SiO_4$: $Cr^{3+}$, $Mn^{2+}$ with the trap depth of 1.05 eV, fulfilling the requirements for optical information storage, was fabricated for the first time through the solution combustion method and subsequent heat treatment at 1000℃ for 2h. The crystal structure, micromorphology, photoluminescence (PL), photoluminescence excitation (PLE), and afterglow properties of $Zn_2SiO_4$: $xCr^{3+}$, $yMn^{2+}$ were studied systematically. By applying the strategy of trap depth engineering, high trap density with proper trap depth was observed when $Cr^{3+}$ ions were introduced into $Zn_2SiO_4$: $Mn^{2+}$. Thermoluminescence (TL) glow-curve analysis through the initial rise (IR) method was conducted to gain some insight into the information of traps. As proof of application, information storage was experimentally achieved by choosing 275 nm illumination for "information writing" and 980 nm NIR excitation for "information reading". The results indicate that $Zn_2SiO_4$: $Cr^{3+}$, $Mn^{2+}$ phosphor holds promise for potential applications in the field of optical information storage.

# 1. Introduction

Information storage safety plays a key role in modern life where so many message steal events may happen every day when the era of big data comes. Therefore, developing secure and more convenient storage media is of great importance to meet the urgent demand [1]. Persistent luminescence materials have been widely investigated in recent years due to the unique excitation-emission process and promising multidimensional anti-counterfeiting recognition such as light wavelength, emission intensity duration time, and so on [2, 3]. Generally speaking, this type of material can be stimulated by light of various wavelengths spanning from X-ray to the visible spectrum. After the light source ceases, electrons stored in traps will be slowly released to the conduction band or some excitation energy levels of the luminescent center under room temperature conditions. Then electrons will transit back to the ground state to emit the sustaining light and finally be exhausted in the trap [2]. Furthermore, deep trap materials are similar to the former, but their traps are deeper (about more than 0.7 eV, but preferably not exceeding 1.2 eV). In the absence of higher-energy light excitation or increased thermal stimulation, electrons will be trapped within these traps and remain preserved at room temperature [4, 5]. According to this principle, optical information storage can be achieved by utilizing these materials [1, 6-8].

Since the persistent luminescent materials $SrAl_2O_4$: $Dy^{3+}$, $Eu^{2+}$ came out, which emits greatly intense green luminescent with the tens of hour afterglow, persistent luminescent materials research has appeared explosive growth [9]. Usually, there exists a competitive relationship between luminescent and afterglow. The host material, doping cation species, intrinsic defects, and even the synthesis environment can significantly affect the luminescence intensity and trap energy levels. These factors could also either hinder or enhance electronic storage capabilities [10, 11]. However, generally, irrespective of the modulation technique used to improve long-lasting luminescence abilities, the key is to adjust the depth of the trap and its corresponding concentration [12]. For example, dopants are a typical tuning method for forming or optimizing traps. J. Du et al explore the $SrSn_2O_4$: $Sm^{3+}$, $Si^{4+}$ with the orange afterglow by doping the $Si^{4+}$ ions to substitute the $Sn^{3+}$ site. The addition of $Si^{4+}$ induces a thermoluminescent peak appears at 170℃, in contrast to the absence of peaks in the undoped matrix [13], endowing the rewriting and reading ability for the system. In addition, Lin et al synthesized the $SrAl_{12}O_{19}$: 0.5%$Mn^{2+}$, 5%$Gd^{3+}$ using solid state reaction method by co-doping $Mn^{2+}$ and $Gd^{3+}$. The materials feature deep traps with multiple energy levels ranging from 0.6 eV to 0.9 eV, rending it responsive to the stimulation of 980 nm, 1064 nm and 1550 nm after illuminating cessation. Notably, the information saved in the materials is innovatively written onto the optical coating disc, depending on the disc's rotational speed and the written-in areas, information can be easily retrieved [14].

Moreover, traps can be modulated by controlling the matrix material itself. Taking $Y_3Al_2Ga_3O_{12}$: $0.015Ce^{3+}$, $0.002V^{3+}$ as the example [8], the $V^{3+}$ in this system acts as the trap center while the $Ce^{3+}$ behaves as the emission center. Through the adjustment of the matrix, the energy gap based on the ratio of the $Al^{3+}$ to $Ga^{3+}$ can be enlarged or shortened, resulting in the controlled trap depth from 1.2 eV to 1.6 eV. Another optimizing method is defect control. In $ZnGa_2O_4$: $Cr^{3+}$ or $Bi^{2+}$ system [15], there are plenty of anti-site defects of $Zn^{2+}$ and $Ga^{3+}$, which can offer the defect energy level for electron storage. Likewise, according to some existing reports, some nitrides also possess similar defect energy levels. For example, $Mg_{0.2}Ca_{0.8}AlSiN_3$: 0.2at% $Eu^{2+}$ whose orange afterglow can last for thousands of seconds [12]. When the afterglow disappears, electrons can be excited from the deep trap energy levels formed by intrinsic defects using a 980 nm near infrared light source.

Herein, employing a doping ion strategy, we reported the deep trap materials $Zn_2SiO_4$: $Cr^{3+}$, $Mn^{2+}$ phosphor (hereafter abbreviated as ZSO: $Cr^{3+}$, $Mn^{2+}$), capable of emitting 521 nm green light. Silicate, known as the stable and multi-doped matrix, is often synthesized via solid state reaction at about 1300℃ [16-20]. On the contrary, by utilizing the solution combustion synthesis route, the sample experiences lower temperatures (1000℃ for 2 hours), allowing for the fabrication of non-agglomerated nano ZSO: xCr$^{3+}$, yMn$^{2+}$ (*x*=0-0.04, *y*=0-0.06) materials with high crystallinity. SEM and TEM images show the grain distribution between 30 nm to 150 nm. The XPS pattern confirmed the nearly unchanged valence rise of Mn$^{2+}$ during the heat treatment process. In this system, Mn$^{2+}$ serves as the emission center and Cr$^{3+}$ introduces the deep trap center. Our research focuses on trap distribution and concentration adjustment. In terms of the precise initial rise method, the trap depth can reach the highest 1.05 eV. The application example of ZSO: Cr$^{3+}$, Mn$^{2+}$ exhibited the potential optical information storage capacity with the bright luminescent intensity.

## 2. Experiment sections

### 2.1. Synthesis

The Zn$_2$SiO$_4$: Cr$^{3+}$, Mn$^{2+}$ phosphors were pre-prepared through the solution combustion method at the low temperature of 500℃ [21]. The raw materials include combustion adjuvant fuel urea (99.999%, Aladdin), Zn(NO$_3$)$_2$·xH$_2$O (99.9%, merck), Cr(NO$_3$)$_3$·9H$_2$O (99.9%, Aladdin) and Mn(NO$_3$)$_2$·4H$_2$O (≥98%, Damas-Beta). The nanoscale SiO$_2$ (99.5%, Picasso) and Tetraethyl orthosilicate (TEOS, 38%~43% SiO$_2$, Aladdin) were also selected as the raw material to explore the effects of silicon sources on the optical properties of samples.

Firstly, all metal nitrates and fuels were weighed according to the stoichiometric ratio, then the mixtures were thoroughly dissolved and sonicated in the deionized water. Subsequently, the solution was stirred at 70℃ and 300 rpm for 15 min until the solution turned into a dark grey gel, which was then transferred into a crucible and ignited in a furnace at 500℃ [22-24]. After the solution combustion was quickly completed, the porous products were collected and ground into powder [25]. Finally, the phosphors were obtained after the heat treatment at 1000℃ for 2 h. The complete synthetic reaction equation for Zn$_2$SiO$_4$: Cr$^{3+}$, Mn$^{2+}$ is given below.

$$2Zn(NO_3)_2 + SiO_2 + mCH_4N_2O + nO_2 \rightarrow Zn_2SiO_4 + (2+m)2N_2(NO/NO_2) + 2mH_2O + mCO_2$$

In addition, the Zn$_2$SiO$_4$: Cr$^{3+}$, Mn$^{2+}$ phosphors were also synthesized by solid state reaction with raw materials of ZnO (99.9%, Aladdin), SiO$_2$ (99.99%, Aladdin),

Cr$_2$O$_3$ (99.95%, Aladdin) and MnO (99.5%, Aladdin) for comparison. The raw materials were thoroughly mixed with the added alcohol. Afterwards, the fried mixtures were calcined in a muffle furnace at 1350°C for 4 h to obtain the final products. Relevant measurement results synthesized by solid-state reaction method were given in Fig. S2.

**2.2. Characterization**

The phase and crystal structure of the ZSO: Cr$^{3+}$, Mn$^{2+}$ phosphors were measured by X-ray diffraction (XRD) with a Bruker D8 Advance X-ray diffractometer (Bruker, Germany) whose radiation source comes from Cu Kα (λ=1.5406 Å) at 40 mA and 40 kV and the diffraction angles were ranging from 10º to 80º with the scan speed of 0.5º/s. The morphology was measured by Scanning electron microscope (SEM, S-4800, Hitachi, Japan). The particle element distribution was performed using an Energy dispersive Spectroscope equipped on the High resolution transmission electron microscope (Tecnai G2 F20, FEI, America). The photoluminescence (PL), photoluminescence excitation (PLE) spectra and persistent luminescence were measured by fluorescence spectrophotometer (F-7000, Hitachi, Japan) at the speed of 240 nm/second with the 500W Xe-lamp and an external 980 nm laser source. The thermoluminescence (TL) spectra were performed with the thermoluminescence dosimeter (FJ-427A, China Nuclear Control System Engineering, China), every sample would be excited for 1 minute under 275 nm light before the beginning of test. X-ray photoelectron spectroscope analysis was carried out by (Thermo escalab 250XI, America). Photos of the appearance and afterglow luminescence of the prepared materials taken by a smartphone camera (iphone12, USA).

**3. Result and discussion**

The space group of Zn$_2$SiO$_4$ is $R\bar{3}$ and belongs to the rhombohedral crystal system. It can be clearly seen in Fig. 1 that the Si$^{4+}$ ions show tetrahedral coordination surrounded by four O$^{2-}$ ions, and Zn$^{2+}$ ions exhibit the same ligand coordination as Si$^{4+}$. Considering the factor of ionic radius that Si$^{4+}$ ($r$=0.2 Å) has smaller ionic radius than that of Zn$^{2+}$ ($r$=0.74 Å), Cr$^{3+}$ ($r$=0.615 Å) and Mn$^{2+}$ ($r$=0.66 Å) dopant ions prefer to occupy the $Zn^{2+}$ sites in $Zn_2SiO_4$ [26, 27]. The calculated lattice parameters based on the results of Rietveld refinement for $Zn_2SiO_4$: $0.005Cr^{3+}$, $0.003Mn^{2+}$ (Fig. S1) are a=13.9455 Å, b=13.9455 Å, c=9.3128 Å, α=90º, β=90º, and γ=120º.

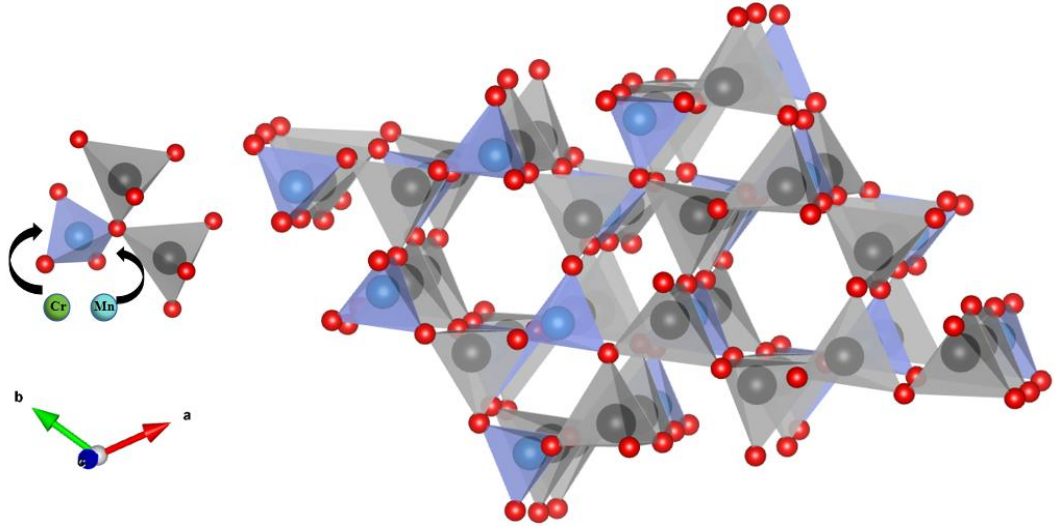

**Fig. 1.** Crystal structure of $Zn_2SiO_4$ (black spheres represent $Si^{4+}$ ions, blue spheres for $Zn^{2+}$ ions and red spheres for $O^{2-}$ ions).

The XRD patterns of ZSO: $0.003Cr^{3+}$, $0.005Mn^{2+}$ sample synthesized via solution combustion method using nanoscale $SiO_2$ and TEOS as the silicon sources are depicted in Fig. 2(a). It's evident that samples using nanoscale $SiO_2$ as the silicon source demonstrate a pure phase, which consistent well with $Zn_2SiO_4$ (JCPDS #79-2005), demonstrating that the little doped $Cr^{3+}$ and $Mn^{2+}$ ions will not change crystal structure. However, there is obvious impurity phase observed in the samples using TEOS as the silicon source, which is consistent with ZnO (JCPDS #75-0576). Therefore, all the samples discussed in this paper were synthesized using nanoscale $SiO_2$.

Fig. 2(b) shows the morphology of ZSO: $0.003Cr^{3+}$, $0.005Mn^{2+}$ synthesized by solution combustion method, representing a typical sample heat-treated at 1000℃ for 2 h. It reveals the uniform grain shape in particle size ranging distinctly from 30 nm to 150 nm around. Fig. 2(c) exhibits the same stochiometric ZSO: $0.003Cr^{3+}$, $0.005Mn^{2+}$ synthesized by solid state reaction. It can be observed that the prepared particles exhibit an equiaxed shape with a diameter of ~ 5 μm.

The crystal structures of the samples were confirmed by using selected area electron diffraction (SAED) patterns based on the high-resolution transmission electron microscopy (HR-TEM). The well dispersive nano materials of $Zn_2SiO_4$: $Cr^{3+}$, $Mn^{2+}$ synthesized by solution combustion method are characterized by TEM image as exhibited in Fig. 3(a). According to the standard unit cell parameters provided by the standard card (JCPDS#79-2005), Fig. 3(b) shows the (3 0 0) crystal plane of the $Zn_2SiO_4$ unit cell, and its inter-plane spacing is 0.3930 nm. In order to confirm the incorporation of $Mn^{2+}$ and $Cr^{3+}$ ions into the host, the EDS mapping was carried out in the selected area in Fig. 3(c). As shown in Fig. 3(d) to Fig. 3(h), Zn, Si, O, Cr and Mn elements are uniformly distributed in the selected area.

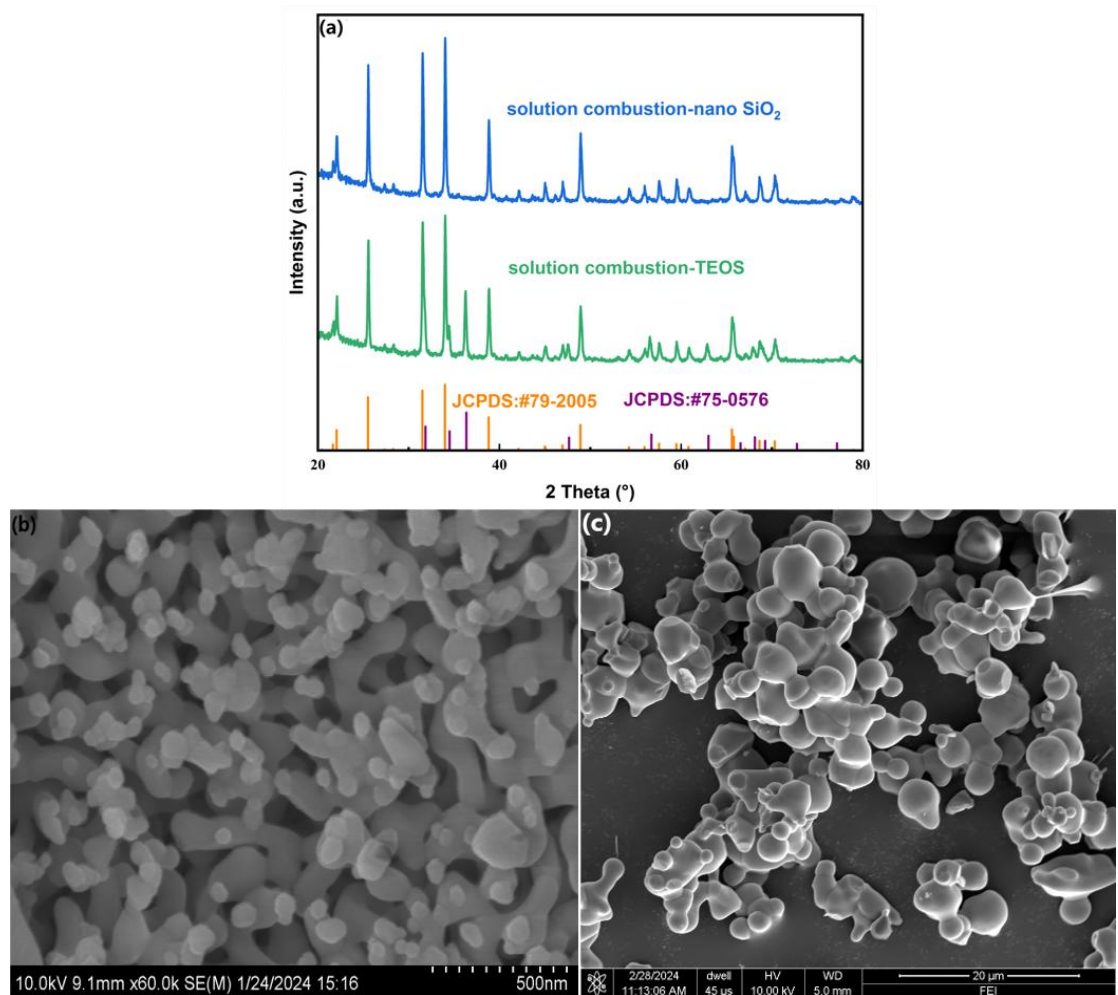

**Fig. 2.** (a) XRD patterns of $Zn_2SiO_4$: $Cr^{3+}$, $Mn^{2+}$ fabricated with different raw materials at 1000℃; the SEM photograph of the typical $Zn_2SiO_4$: $0.003Cr^{3+}$, $0.005Mn^{2+}$ synthesized by solution combustion (b) and solid state reaction (c).

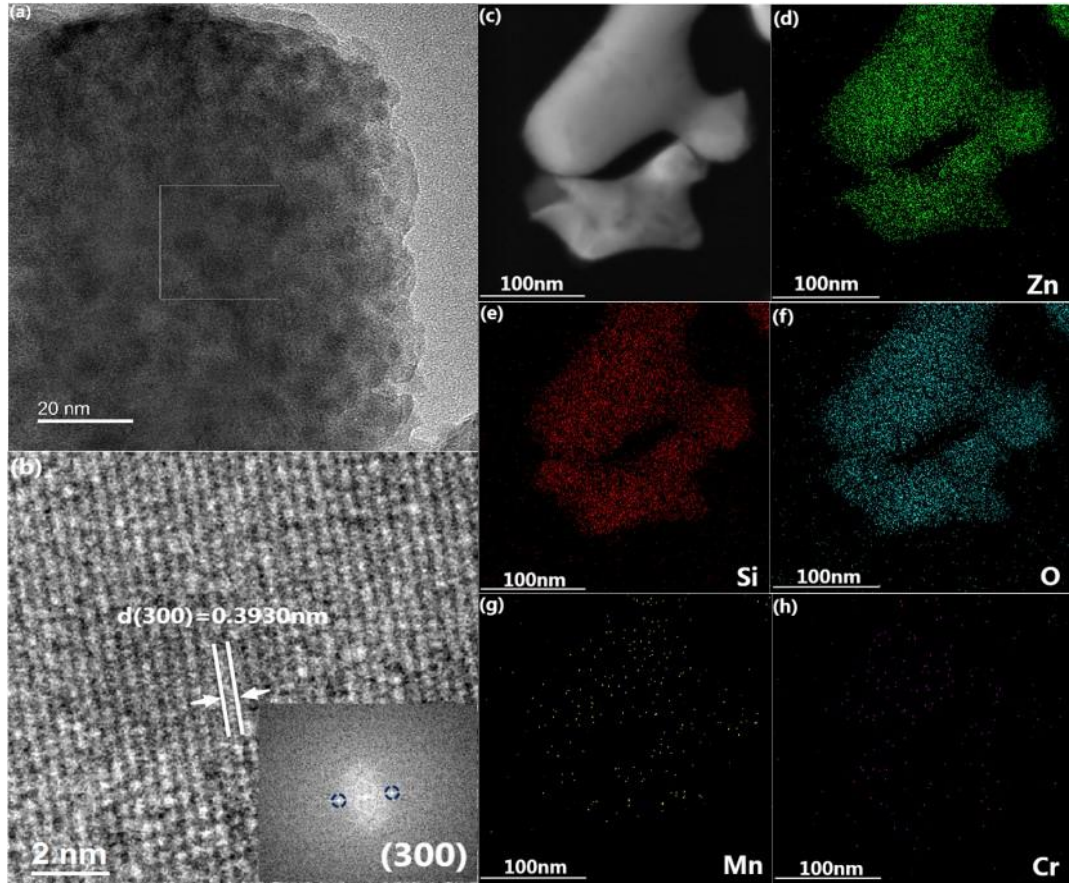

Fig. 3. (a) TEM image of $Zn_2SiO_4$: $Cr^{3+}$, $Mn^{2+}$ and the square is the selected diffraction fringe area in Fig. 3(b). (b) HR-TEM image and the inset is obtained by Fourier transform. (c) to (h) Elemental distribution mapping of ZSO: $Cr^{3+}$, $Mn^{2+}$ which is equipped on HRTEM.

In general, the materials synthesized by doping with $Mn^{2+}$ ions usually show a shift in valence of Mn element from divalent to trivalent or even tetravalent [28, 29], which may be related to the synthesis conditions when doping with manganese ions. Herein, this work includes a heat treatment at 1000°C, which may lead to lead to an increase in the valence state of $Mn^{2+}$ ions. Therefore, the valence of Mn elements of the sample were studied by X-ray photoelectron spectroscopy (XPS) analysis.

First of all, it is worth mentioning that the binding energy signal of the $Mn^{2+}$ ion in the $Zn_2SiO_4$ system will be interfered by the Auger electron signal of the $Zn^{2+}$ ions, which is approximately around 652 eV of the 2p orbital of the Mn. Therefore, the uninterrupted energy spectrum signal before 650 eV is used to fit and analyze the valence state components. In Fig. 4, The fitted curves can be roughly divided into 11, of which 6 blue curves correspond to the peak positions of $Mn^{2+}$ ions, and 5 orange curves correspond to the peak positions of $Mn^{3+}$ ions [30]. Blue peaks located at 640 eV, 641.22 eV, 642.23 eV, 642.98 eV, 647.51 eV and 644.84 eV occupied the main peak area with the proportion of 83.53%. Among those, the shake-up peak located at 647.51eV was the characteristic peak distinguishing $Mn^{2+}$ from $Mn^{3+}$ and $Mn^{4+}$. Returning to the curve of $Mn^{3+}$ ions, the peak positions of $Mn^{3+}$ ions can be fitted at 641.24 eV, 641.94 eV, 642.76 eV, 643.7 eV and 645.08 eV, and their area only accounts for about 16.47%. The ratio of $Mn^{2+}$ to $Mn^{3+}$ ions is approximately 5:1, thus explaining the almost unchanged valence state during synthesis and heat treatment.

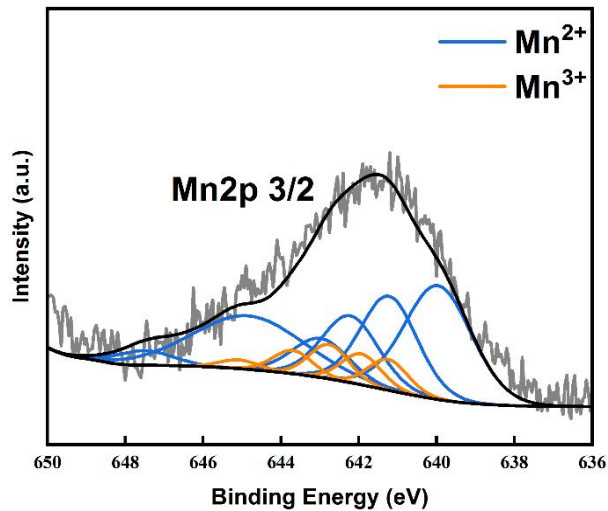

**Fig. 4.** XPS spectra and the fitting curves of $Mn2p_{3/2}$ orbit based on ZSO: $Cr^{3+}$, $Mn^{2+}$.

Fig. 5(a) shows the photoluminescence spectra (PL) and photoluminescence excitation spectra (PLE) of the ZSO: $0.01Cr^{3+}$, $xMn^{2+}$, ($x$=0.01-0.06) phosphors. By monitoring the emission at 521 nm, all the phosphors show two intense absorption peaks in the range from 200 nm to 300 nm. The peak centered at 251 nm is ascribed to the electron transition from $^6A_1$ ($^6S$) to $^4T_1$ ($^4G$) in the $Mn^{2+}$ energy level. The other peak centered at 212 nm can be regarded as the intrinsic absorption of the $Zn_2SiO_4$ host, in which the electrons transition from the valence band to the conduction band. It is also reported in the literatures that the host absorption peak ranges from approximately 205 nm and 225 nm, which is in high agreement with the present results [31-33]. With the doping concentration of $Mn^{2+}$ increasing from 0.01 to 0.06, the absorption peak located near 251 nm gradually shifts from 243 nm to 251 nm, and its luminescent intensity firstly increases and then decreases with the concentration quenching value of $x$=0.05. However, the other absorption peak located near 212 nm shows the opposite blue shift, and its luminescent intensity tends to decrease. The emission spectra of ZSO: $0.01Cr^{3+}$, $xMn^{2+}$ ($x$=0-0.06) phosphors under 251 nm excitation are shown in Fig. 5(a). The only symmetrical emission peak is located near 521 nm, which is ascribed to the $^4T_1(^4G) \rightarrow {}^6A_1(^6S)$ electronic transition of the $Mn^{2+}$ ions on the substituted $Zn^{2+}$ position. As the doping concentration of $Mn^{2+}$ ions increases, the emission peaks are red-shifted due to the relatively sensitive crystal field effect of $Mn^{2+}$ ions. Since the inherently high spin state of $Mn^{2+}$ tends to shorten its electronic transition energy level located between the ground state and the excited state, thus results in a shift of the peak position towards longer wavelengths [34].

The persistent luminescent decay curves of the ZSO: $0.01Cr^{3+}$, $xMn^{2+}$ ($x$=0.01, 0.04, 0.06) phosphors after 251 nm irradiation for 5 min are exhibited in Fig. 5(b). After the irradiation, the afterglow intensity of all the samples decreases rapidly with time. And with the increasing doping concentration of $Mn^{2+}$, the initial afterglow intensity gradually decreases, the decay is faster as well as the persistent luminescence performance becomes worse.

As is known to all, the optical materials with best persistent luminescence require appropriate trap distribution. Commonly, thermoluminescence (TL) intensity can be considered as the thermoluminescence of charge electrons released from the trap energy levels. Before the TL test, all the samples were irradiated for 1 min under 275 nm UV source. To investigate the influence of $Mn^{2+}$ and $Cr^{3+}$ ions doping on the depth and concentration of trap of ZSO: $0.01Cr^{3+}$, $xMn^{2+}$, a series of TL spectra of ZSO: $Cr^{3+}$, $xMn^{2+}$ were tested to investigate the optimum doping concentration for the ZSO: $Cr^{3+}$, $xMn^{2+}$.

The TL glow curves of the ZSO: ZSO: $0.01Cr^{3+}$, $xMn^{2+}$ ($x$=0.01-0.06) phosphors were depicted in Fig. 5(c). All the samples exhibit a narrow peak1 ranging from 50℃ to 120℃ with relative high intensity whose trap depth is around 0.729 eV in terms of the empirical formula $E=T_m/500$ where $T_m$ represents the peak temperature. Another wide peak2 is ranging from 150℃ to 275℃ with the much lower intensity than peak1. With the $Mn^{2+}$ ion doping concentration increasing from 0.01 to 0.06, the position of

peak1 gradually moves monotonically from 91℃ to 87℃. In addition, the TL glow intensity of both peak1 and peak2 decreases monotonously with the $Mn^{2+}$ concentration increasing. The difference between these two peaks is that the decrease in peak2 is faster. When $x$=0.02, the intensity of TL peak2 almost disappears, while the intensity of TL peak1 descends at a relatively slow rate. Therefore, it is reasonable to speculate that the high concentration of $Mn^{2+}$ ions will not only suppress the trap concentration, but also lead to a shift of the trap depth towards the low-temperature part.

Similarly, the TL glow spectra of ZSO: $y$$Cr^{3+}$, 0.01$Mn^{2+}$ ($y$=0-0.04) are showed in Fig. 5(d) and the PL and PLE spectra were showed in Fig. S3. In Fig. 5(d), the TL glow curves of ZSO: $y$$Cr^{3+}$, 0.01$Mn^{2+}$ ($y$=0-0.04) show two peak positions which is identical with the Fig. 5(c). Differently, with the concentration of $Cr^{3+}$ ions increasing, the TL peak1 moves from 91℃ to 98℃. The intensity of peak1 rises slowly until $y$=0.005 and then to decrease sharply, which means that the trap concentration also begins to decrease when $y$ exceeds 0.007. As for peak2, the trap depth of TL peak2 is approximately 0.946 eV (200℃). When $y$=0.001, the intensity of TL peak2 is greatly suppressed. With a further increase in the concentration of $Cr^{3+}$ ions, the intensity of TL peak2 is almost is so weak to be almost unobservable. Therefore, considering that $Cr^{3+}$ ions will weaken the luminescent intensity of the material and the luminescent ability is also an important factor for deep trap materials, so the value of $y$=0.003 is regarded as the optimal doping concentration of $Cr^{3+}$.

Based on the above results analysis, it can be concluded that the change process of trap distribution after doping with $Mn^{2+}$ and $Cr^{3+}$ ions in $Zn_2SiO_4$. In this system, $Mn^{2+}$ ions doping will provide two kinds of traps for electrons storage, which correspond to peak1 and peak2 mentioned above. However, when the doping concentration of $Mn^{2+}$ exceeds 0.01, the concentration of trap1 will decrease gradually while the trap2 will directly disappear. Meanwhile, trap1 will move toward low temperature slowly with the increasing concentrations of $Mn^{2+}$. Differently, $Cr^{3+}$ ions only provide one trap. With the concentration of $Cr^{3+}$ ions increasing, the concentration of trap1 will rise rapidly and then decrease, while trap2 will disappear directly. Contrary to the influence of $Mn^{2+}$, the peak1 position moves towards high temperature part with the concentration of $Cr^{3+}$ ions increasing. To clearly analyze the regulation of TL curves, the TL glow curve of undoped $Zn_2SiO_4$ is also showed in Fig. S4 for comparison, $Cr^{3+}$ behaves more efficiently on the concentration of trap1 than $Mn^{2+}$ on that. What's more, $Cr^{3+}$ usually acts as the main trap center in plenty of materials matrix such as the classical system $Y_3Al_2Ga_3O_{12}$: $Ce^{3+}$, $Cr^{3+}$ [35, 36]. Therefore, the trap1 is dominated by $Cr^{3+}$ while $Mn^{2+}$ make contribution to the trap1 and trap2.

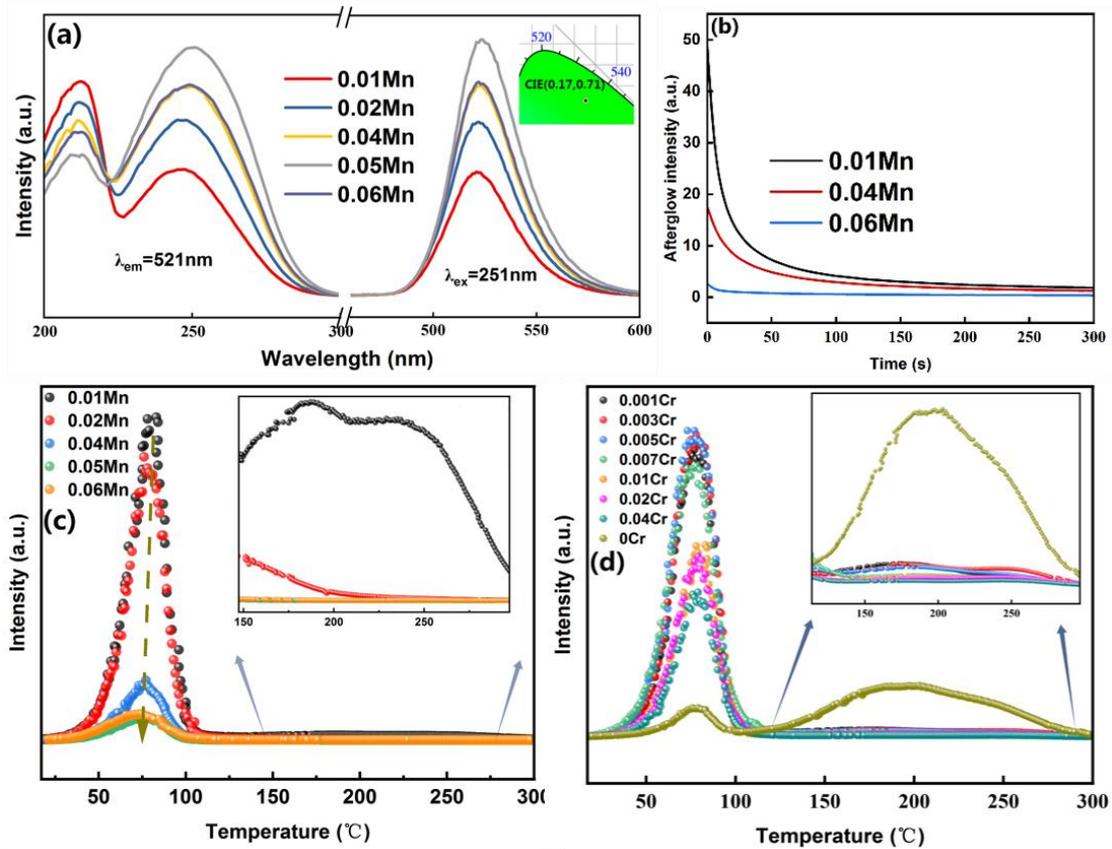

**Fig. 5.** **(a)** Photoluminescent excitation (PLE) and photoluminescent (PL) spectra of ZSO: $0.01Cr^{3+}$, $xMn^{2+}$ ($x$=0.01~0.06); **(b)** Afterglow decay curves of ZSO: $0.01Cr^{3+}$, $xMn^{2+}$ ($x$=0.01, 0.04, 0.06) monitored at 521 nm emission under 251 nm excitation; **(c)** Thermoluminescence (TL) glow spectra of ZSO: $0.01Cr^{3+}$, $xMn^{2+}$ ($x$=0.01-0.06); **(d)** TL glow spectra of ZSO: $yCr^{3+}$, $0.01Mn^{2+}$ ($y$=0-0.04).

Based on the optimum concentration of $Cr^{3+}$ (0.003), a series of ZSO phosphors doped with various content of $Mn^{2+}$ were systematically studied. The TL glow curves of the ZSO: $0.003Cr^{3+}$, $xMn^{2+}$ ($x$=0.001-0.01) are exhibited in Fig. 6. When the $Mn^{2+}$ ion doping is 0.005, the TL intensity of peak1 reaches the maximum. It is worth mentioning that the movement of peak1 towards the low temperature region is not obvious, which ensures relatively deep traps for storing electrons. When $x$ exceeds 0.005, the TL intensity of peak1 commences descending. During the whole increasing process of the content of $Mn^{2+}$, the intensity of peak1 falls to show the matched contribution on the concentration of trap1 compared with the counterpart of $Cr^{3+}$ in Fig. 5(d). Furthermore, with the concentration of $Mn^{2+}$ ions increasing, the TL intensity of peak2 can always be observed. Utilizing trap2, more electrons will be less likely to be excited at room temperature, so that these electrons can be stored in trap2 for a long time. In addition, the inset in Fig. 6 illustrates the monotonous rise of PL intensity with the concentration of $Mn^{2+}$ increasing and the corresponding PL spectra are displayed in Fig. S5. ZSO: $0.003Cr^{3+}$, $0.005Mn^{2+}$ phosphor ensures excellent luminescent intensity and it is exactly at the turning point of the slope. As the $Mn^{2+}$ concentration exceeds 0.005, the slope growth slows down.

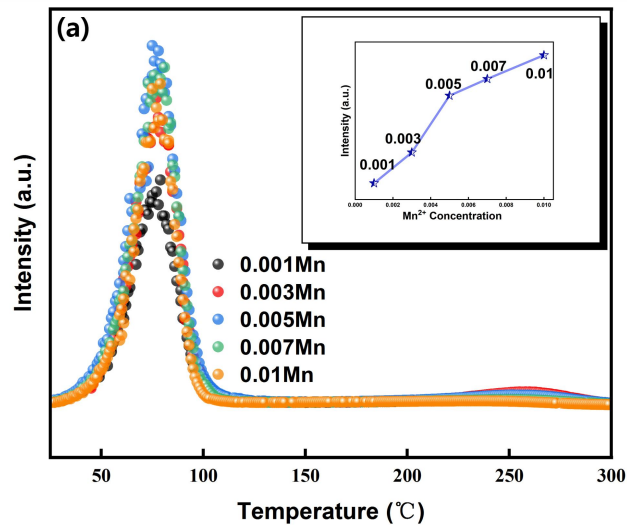

**Fig. 6.** Thermoluminescence spectra curves of ZSO: $0.003Cr^{3+}$, $xMn^{2+}$ ($x$=0.001~0.01) (inset is the photoluminescent intensity monitored at 521nm under 251nm excitation source with varying $Mn^{2+}$ concentrations).

Overall, the phosphor ZSO: $0.003Cr^{3+}$, $0.005Mn^{2+}$ with the optimum persistent luminescence was selected to explore the accurate trap distribution with the $T_m$-$T_{stop}$ method [37]. First of all, the phosphor was excited by 275 nm UV source for 1 min at room temperature, which ensures the fully occupied trap states by electrons. Secondly, the excited phosphor was heated to the setting temperature $T_{stop}$ with the heating speed of 1 K/s, then the heated phosphor was naturally cooled down to room temperature. Notably, the second process including heating and cooling was kept for the same time. Finally, the phosphor was reheated with the TL spectra recorded.

Herein, the TL glow spectra of the ZSO: $0.003Cr^{3+}$, $0.005Mn^{2+}$ with different $T_{stop}$ temperatures ranging from 298 K to 388 K are depicted in Fig.7(a). As the $T_{stop}$ increases, the asymmetric TL curves gradually shifts towards the high temperature side and the intensity decreases. These processes of change illustrate that the dynamics of trap and de-trap process in ZSO: $0.003Cr^{3+}$, $0.005Mn^{2+}$ belongs to neither the first order nor the second order, since the indistinct overlapping peaks also shift to higher temperature. Therefore, suitable analysis methods are needed to exclude the effects of dynamic order and overlapping peaks.

Conventionally, there are many methods to analyze the trap distribution of long persistent phosphors such as the peak shape method [38, 39], computer fitting method [40], peak position method etc., among which, the initial rise analysis method possesses the merits of practical physical significance in results, non-empirical calculation process. Importantly, it is irrelevant with the dynamic order and overlapped peaks during the rising process at low temperature. Therefore, the method is suitable for analyzing the trap distribution of ZSO: $0.003Cr^{3+}$, $0.005Mn^{2+}$.

The TL glow curves in the initial rise part can be semi-quantitatively described by formula $I(T)= Cexp(-E/kT)$, where I is the TL glow intensity function with temperature varying, C refers to a constant representing frequency index, k means the Boltzmann's constant and T is the temperature. By this formula, the trap depth E is related to the fitting slope in the initial rising temperature stage [41], and the trap concentration is decided by integral area difference of two adjacent curves [8].

According to the formula, the TL glow curves in Fig. 7(a) is plotted as the function of ln(I) against 1000/T in Fig. 7(b). Then the trap depth (E) and the trap concentration can be calculated in terms of the initial rise method. In the low temperature region, the fitting slope of the curve is supposed to be linear. Based on the theory, the trap depth (E) was calculated from that part, and the distribution of trap depth (E) with $T_{stop}$ increasing was depicted in Fig. 7(d). At the same time, the normalized distribution of the trap density is depicted in Fig. 7(c), which exhibits the relatively precise estimated results according to the integral area difference mentioned above.

As shown in Fig. 7(c), it can be clearly distinguished that there are two kinds of trap distribution. The one is mainly distributed in the range from 0.8 eV to 0.95 eV while the other one is distributed in the range from 0.95 eV to 1.05 eV. For first trap distribution, it appears relative less normalized density, which can be explained to be the influence of $Mn^{2+}$ because of the shift of TL peak toward low temperature in Fig. 5(c). In addition, the afterglow monitored at 521 nm in Fig. 5(b) can also be ascribed to be derived from this part considering the low trap density between 0.8 eV and 0.95 eV. For the other one, which accounts for dominating density ranging from 0.95 eV to 1.05 eV, it can be attributed to the influence of $Cr^{3+}$, for the reason that $Cr^{3+}$ do more influence on TL intensity than $Mn^{2+}$ and the TL peak moves towards high temperature in Fig. 6(d).

The fitting slope distribution as a function of trap depth against $T_{stop}$ is showed in Fig. 7(d). It can be apparently divided into two straight lines with different slopes, which conveys that there exist at least two kinds of traps in ZSO: $0.003Cr^{3+}$, $0.005Mn^{2+}$. Considering $Mn^{2+}$ and $Cr^{3+}$ have the capacity to influence the trap concentration, the relative deep trap is attributed to the $Cr^{3+}$ while the shallower trap is assigned to the $Mn^{2+}$.

Generally, the trap depth of 0.7 eV to 1.2 eV is suitable for electron storage which is not readily escaped from the traps in room temperature. Deeper trap depth may cause the dilemma of electron escape from traps while shallower trap depth is the reason of persistent luminescence whose trap depth is around 0.4 eV to 0.7 eV [42]. To further investigate the luminescent mechanism and propose the application of optical information storage for this materials, a mask model was used to explore the feasibility.

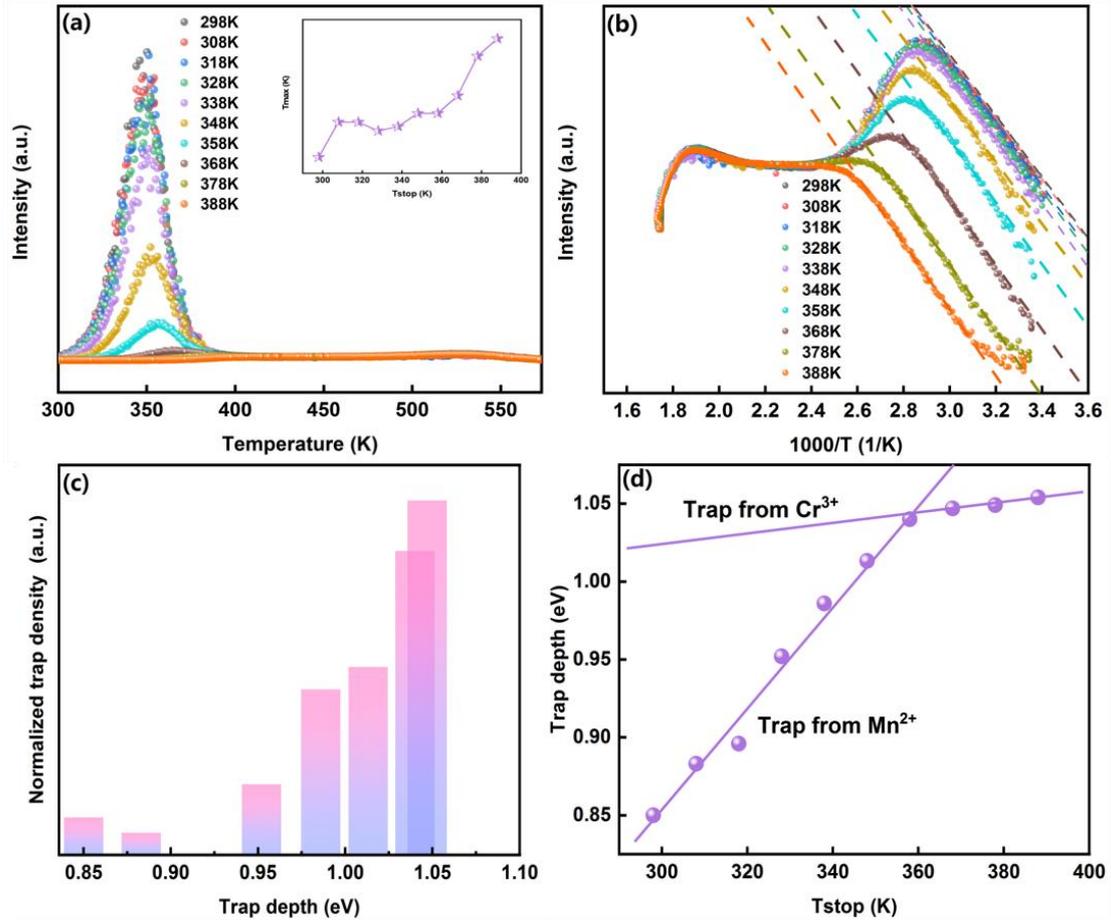

**Fig. 7.** (a) TL glow spectra of ZSO: $0.003Cr^{3+}$, $0.005Mn^{2+}$ at different temperature from 298K to 388K, both the preliminary excitation under 275nm light source and the preheating process time are controlled for the same time towards each sample, and the inset in the image is the relationship between preheating temperature ($T_{stop}$) and the respective peak position for each curve. (b) The fitting curves of ZSO: $0.003Cr^{3+}$, $0.005Mn^{2+}$ based on the initial rise analysis (c) the normalized trap distribution of ZSO: $0.003Cr^{3+}$, $0.005Mn^{2+}$ according to the integral area. (d) The calculated results of trap depth which change with the preheating temperature ($T_{stop}$).

Based on the analysis of trap distribution for the application of optical information storage, the relevant measurements of afterglow were conducted to verify that. As showed in Fig. 8(a), when using an excitation light source of 251 nm, the phosphor ZSO: $0.003Cr^{3+}$, $0.005Mn^{2+}$ show the stable luminescent intensity within the first 70 s. However, when the light source was removed, the luminescent intensity drops rapidly to 0, which is because the traps of the ZSO: $0.003Cr^{3+}$, $0.005Mn^{2+}$ are relatively concentrated, only a small number of electrons will escape through afterglow. Therefore, such an afterglow decay curve confirms that the phosphors provide long-term stable electron storage capabilities for optical information storage, that is, without a large number of electrons spontaneously escaping from the traps at room temperature. At the same time, the written information will not be displayed due to the afterglow, and the security of information storage is also ensured.

The repeatable writing and reading application can be realized by irradiation of 980 nm near-infrared laser. The impulse shape decay curves of ZSO: $0.003Cr^{3+}$, $0.005Mn^{2+}$ are depicted in Fig. 8(b). The phosphor was excited for 300 s under 251 nm irradiation firstly representing step1 in Fig. 8(b). After the cessation of 251 nm stimulation, the shutter was closed for 60 s (step2). In the step2, the afterglow intensity showed a decreasing trend at first and then became stable immediately, since only a small number of electrons were excited under the thermal disturbance at RT, and more electrons would be trapped in the deep traps. Subsequently, the 980 nm NIR laser (0.5W) was used to release the electrons stored in the deep traps, corresponding to the consumption of electrons in the deep traps between 60s and 120s in Fig. 8(b). When the 980 nm NIR laser was turned off, the photo-stimulated persistent luminescence disappeared, because there was not enough energy to release the electrons from the deep traps, corresponding to the part from 120 s to 180 s. The circulations were carried out for 10 times, and each period appeared the same shape of luminescence intensity that excited parts decreased slowly while unexcited parts behaved as a straight line with the time elapsing.

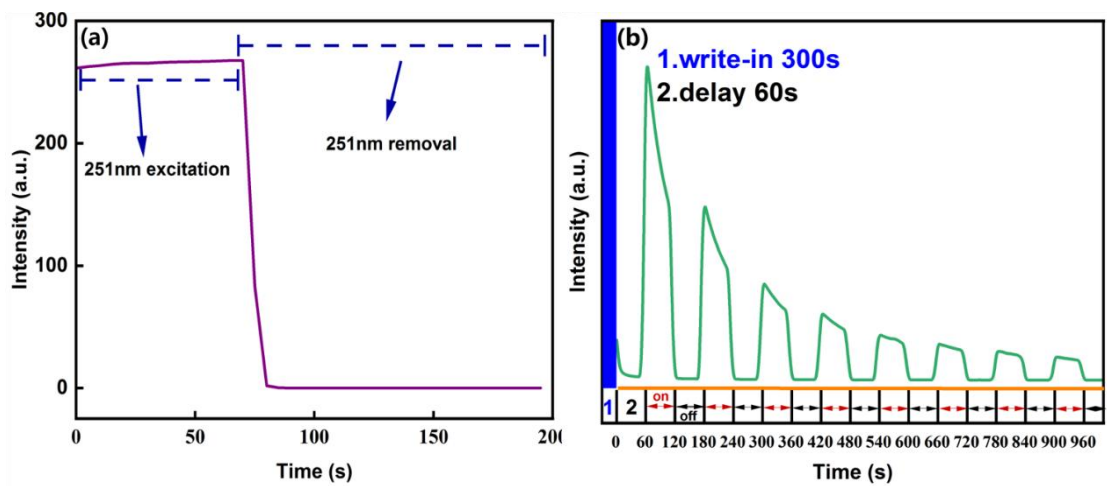

**Fig. 8.** (a) The persistent luminescent decay curve of ZSO: $0.003Cr^{3+}$, $0.005Mn^{2+}$ sample monitored at 521 nm. (b) The decay curves of ZSO: $0.003Cr^{3+}$, $0.005Mn^{2+}$ monitored at 521 nm with the 980 nm NIR laser periodic excitation, and samples were stimulated in the first for 300 s with the 251 nm source.

Application for repeated writing-read function has been realized based on the potential electron storage capacity of ZSO: $0.003Cr^{3+}$, $0.005Mn^{2+}$. A series of illustration images were taken under the dark condition and showed in the following pictures. From Fig. 9(a) to Fig. 9(d), the illustrations of afterglow decay were exhibited while write-in and read-out processes were showed from Fig. 9(e) to Fig. 9(g).

First, the ZSO: $0.003Cr^{3+}$, $0.005Mn^{2+}$ was pressed into a pill (Fig. S6). Next, with utilization of the stimulation of 275 nm for 1 min, the dazzling green emission is showed in Fig. 9(a). After the removal of 275 nm, the afterglow intensity dropped sharply within the first 1s, but there is still weak green emission, as showed in Fig. 9(b). When the time came to 3 s and 5 s, the emission disappeared. The images in Fig. 9(c) and Fig. 9(d) were the process of afterglow change caused by lower and shallower trap concentration, which was strongly supported the decay curves of ZSO: $0.003Cr^{3+}$, $0.005Mn^{2+}$ in Fig. 8(a).

Information storage materials are often stimulated by 980 nm (1.265eV) NIR source. Therefore, the NIR source of 980 nm was selected to realize the application of optical information storage. Firstly, the charged pill was excited by 980 nm NIR source after the afterglow disappeared. As showed in Fig. 9(e), the visible brightness was excited out, which confirmed that the electrons could be stimulated from the deep trap (0.95 eV-1.05 eV). Secondly, the pill was thermally stimulated to release all the electrons stored in the deep traps, as the Fig. 9(f) displayed. Thirdly, the pill was put on a preliminarily written mask cover to embed the information (RUC pattern in Fig. 9(g)) into it. Then, the pill was illuminated by 275 nm UV source, corresponding to the write-in process. After the removal of 275 nm UV excitation, the pill appeared invisible brightness. Later, when the pill was excited by 980 nm NIR source, the illuminated part was stimulated, and the symbol of RUC appeared, corresponding to the read-out process (Fig. 9(h)). Such a write-in and read-out process was sufficiently reflected by impulse-like decay curve exhibited in Fig. 8(b).

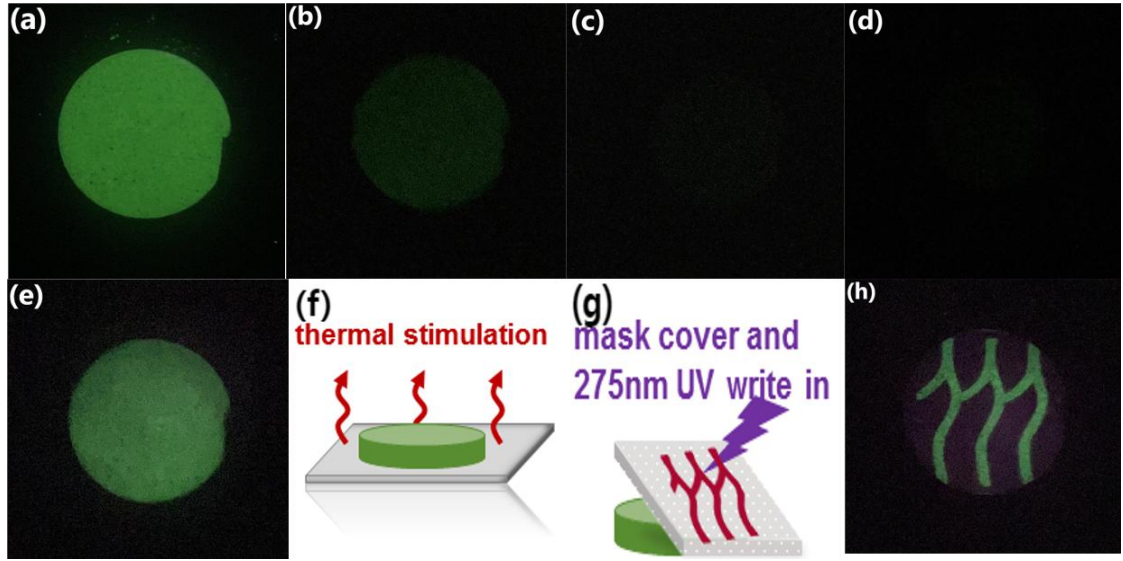

**Fig. 9.** (a) The image of ZSO: $0.003Cr^{3+}$, $0.005Mn^{2+}$ under the 275 nm source illumination. (b) to (d) The schematic diagrams for the afterglow of the ZSO: $0.003Cr^{3+}$, $0.005Mn^{2+}$ at 1s, 3s, 5s without 275 nm excitation. (e) The image of the sample under the excitation of 980 nm NIR source. (f) The schematic diagram of the thermal stimulation process to exhaust all the trapped electrons. (g) The schematic diagram of the write-in process. (h) The images for read-out process of sample with the 980 nm NIR source excitation. All the pictures were taken by smartphone camera with the parameter of ISO6400/ f1.6/ 1/15s shutter time.

Electron storage capacity, especially trap depth, is a critical factor in writing and reading materials. The electrons stored in the traps are susceptible to environmental influences. The schematic diagram of luminescent mechanism of ZSO: $Cr^{3+}$, $Mn^{2+}$ is depicted in Fig. 10, which shows the process including external excitation and internal electrons transfer based on the PL, PerL, TL analysis and experiment phenomenon.

**(1)** With the excitation source involving 251 nm and 212 nm UV light irradiation, electrons in the valence band are excited to the conduction band directly, and electrons on the $Mn^{2+}$ ground state $^6A_1(S)$ transfer to excited state $^4E(G)$ of $Mn^{2+}$. **(2)** Soon after, a small number of activated electrons in the conduction band return to the excited state energy level $^4E(G)$. Then they relax to $^4T_2(G)$ and $^4T_1(G)$ through the nonradiative transition, and transition to $Mn^{2+}$ ground state with the 521 nm emission. Notably, electrons in this process include two kinds of origination, namely electrons transfer from conduction band and electrons transfer from $^6A_1(S)$. However, most electrons will jump into $Cr^{3+}$ traps. **(3)** Therefore, after the UV light source is removed for a long time, using the excitation of the 980 nm NIR light source, the electrons staying at the $Cr^{3+}$ energy level will be released from the traps to the $Mn^{2+}$ excited state energy level, and then relax briefly, with the radiation of photons, causing the appearance of 521 nm green light emission.

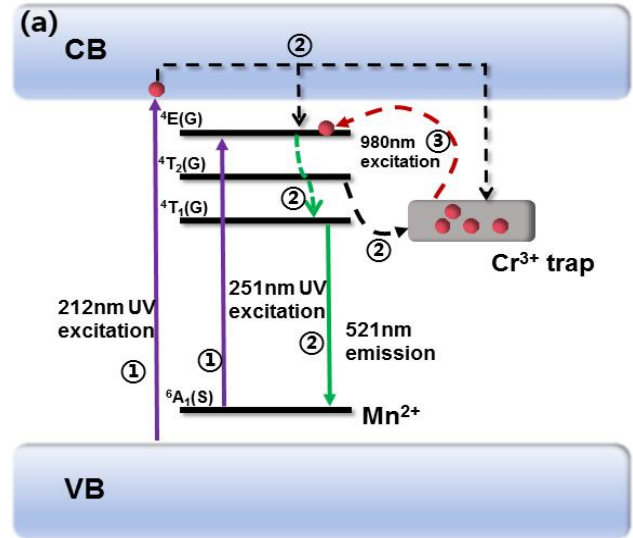

**Fig. 10.** The proposed energy level diagram of ZSO: $Cr^{3+}$, $Mn^{2+}$ phosphor for luminescence and trap-detrap process.

**4. Conclusion.**

In summary, a series of nano phosphors $Zn_2SiO_4$: $yCr^{3+}$, $xMn^{2+}$ ($x$=0~0.6, $y$=0~0.4) were successfully synthesized by efficient solution combustion method with the deepest trap depth of 1.05 eV around to achieve the optical information storage. The phosphors exhibit bright green light peaking at 521 nm around under the 251 nm UV stimulation. In this system, $Mn^{2+}$ acts as emission center with the 521 nm emission attributed to $^4T_1(G) \rightarrow {}^6A_1(S)$ transition under the 251 nm UV radiation while $Cr^{3+}$ behaves as the main trap center. The SEM and TEM has ensured the micro nano structure between 30 nm to 150 nm and the confirmed rhombohedral crystal system. The EDS scanning illustrates the incorporation of $Mn^{2+}$ and $Cr^{3+}$ into the host. Meanwhile, considering the possibility of valence rise, XPS was used to demonstrate the high proportion of $Mn^{2+}$. Through the thermoluminescent analysis, both $Cr^{3+}$ and $Mn^{2+}$ can influence the trap distribution. Through the initial rise method, the accurate traps distribution of the material is between 0.8 eV and 1.05 eV. Among all samples, ZSO: $0.003Cr^{3+}$, $0.005Mn^{2+}$ with the optimum electron storage capacity and hardly seen afterglow was selected to display the desirable application. We do believe it is of great promising to promote the as-synthesized materials to further.

**References**

[1] Y.X. Zhuang, L. Wang, Y. Lv, T.L. Zhou, R.J. Xie, Optical Data Storage and Multicolor Emission Readout on Flexible Films Using Deep-Trap Persistent Luminescence Materials, Advanced Functional Materials 28(8) (2018).

[2] L.F. Yuan, Y.H. Jin, Y. Su, H.Y. Wu, Y.H. Hu, S.H. Yang, Optically Stimulated Luminescence Phosphors: Principles, Applications, and Prospects, Laser Photon. Rev. 14(12) (2020) 34.

[3] Z.W. Long, Y.G. Wen, J.H. Zhou, J.B. Qiu, H. Wu, X.H. Xu, X. Yu, D.C. Zhou, J. Yu, Q. Wang, No-Interference Reading for Optical Information Storage and Ultra-Multiple Anti-Counterfeiting Applications by Designing Targeted Recombination in Charge Carrier Trapping Phosphors, Advanced Optical Materials 7(10) (2019) 11.

[4] J. Xu, S. Tanabe, Persistent luminescence instead of phosphorescence: History, mechanism, and perspective, Journal of Luminescence 205 (2019) 581-620.

[5] K. Van den Eeckhout, A.J.J. Bos, D. Poelman, P.F. Smet, Revealing trap depth distributions in persistent phosphors, Physical Review B 87(4) (2013) 045126.

[6] F. Gao, Q. Pang, D. Gao, C. Jia, H. Xin, Y. Pan, Y. Wang, S. Yun, $Mn^{2+}$-Activated Photostimulable Persistent Nanophosphors by $Pr^{3+}$ Codoping for Rewritable Information Storage, ACS Applied Nano Materials 6(4) (2023) 3054-3064.

[7] C. Liao, H. Wu, H.J. Wu, L.L. Zhang, G.H. Pan, Z.D. Hao, F. Liu, X.J. Wang, J.H. Zhang, Electron Trapping Optical Storage Using A Single-Wavelength Light Source for Both Information Write-In and Read-Out, Laser Photon. Rev. 17(8) (2023).

[8] W. Li, Y. Zhuang, P. Zheng, T.-L. Zhou, J. Xu, J. Ueda, S. Tanabe, L. Wang, R.-J. Xie, Tailoring Trap Depth and Emission Wavelength in $Y_3Al_{5-x}Ga_xO_{12}:Ce^{3+},V^{3+}$ Phosphor-in-Glass Films for Optical Information Storage, ACS Applied Materials & Interfaces 10(32) (2018) 27150-27159.

[9] T. Matsuzawa, Y. Aoki, N. Takeuchi, Y. Murayama, New long phosphorescent phosphor with high brightness, $SrAl_2O_4:Eu^{2+},Dy^{3+}$, Journal of the Electrochemical Society 143(8) (1996) 2670-2673.

[10] F. Clabau, X. Rocquefelte, S. Jobic, P. Deniard, M.H. Whangbo, A. Garcia, T. Le Mercier, Mechanism of phosphorescence appropriate for the long-lasting phosphors $Eu^{2+}$-doped $SrAl_2O_4$ with codopants $Dy^{3+}$ and $B^{3+}$, Chem. Mat. 17(15) (2005) 3904-3912.

[11] R. Hu, Y. Zhang, Y. Zhao, X.S. Wang, G.R. Li, C.Y. Wang, UV-Vis-NIR broadband-photostimulated luminescence of $LiTaO_3:Bi^{3+}$ long-persistent phosphor and the optical storage properties, Chem Eng J 392 (2020) 12.

[12] K.M. Zhu, Z.L. Chen, Y.Z. Wang, H. Liu, Y.D. Niu, X. Yi, Y.H. Wang, X.Y. Yuan, G.H. Liu, $(M,Ca)AlSiN_3:Eu^{2+}$ (M=Sr, Mg) long persistent phosphors prepared by combustion synthesis and applications in displays and optical information storage, Journal of Luminescence 252 (2022) 9.

[13] J. Du, S. Lyu, K. Jiang, D. Huang, J. Li, R. Van Deun, D. Poelman, H. Lin, Deep-level trap formation in Si-substituted $Sr_2SnO_4:Sm^{3+}$ for rewritable optical information storage, Materials Today Chemistry 24 (2022) 100906.

[14] X. Lin, K. Deng, H. Wu, B. Du, B. Viana, Y. Li, Y. Hu, Photon energy conversion and management in $SrAl_{12}O_{19}: Mn^{2+}, Gd^{3+}$ for rewritable optical information

storage, Chem Eng J 420 (2021) 129844.

[15] Y.X. Zhuang, J. Ueda, S. Tanabe, Enhancement of Red Persistent Luminescence in Cr3+-Doped ZnGa2O4 Phosphors by Bi2O3 Codoping, Appl. Phys. Express 6(5) (2013) 4.

[16] J.S. Kim, Y.H. Park, S.M. Kim, J.C. Choi, H.L. Park, Temperature-dependent emission spectra of M2SiO4:Eu2+ (M = Ca, Sr, Ba) phosphors for green and greenish white LEDs, Solid State Commun. 133(7) (2005) 445-448.

[17] X.X. Xu, Q.Y. Shao, L.Q. Yao, Y. Dong, J.Q. Jiang, Highly efficient and thermally stable Cr3+-activated silicate phosphors for broadband near-infrared LED applications, Chem Eng J 383 (2020) 8.

[18] Y.H. Lin, C.W. Nan, X.S. Zhou, J.B. Wu, H.F. Wang, D.P. Chen, S.M. Xu, Preparation and characterization of long afterglow M2MgSi2O7-based (M: Ca, Sr, Ba) photoluminescent phosphors, Mater. Chem. Phys. 82(3) (2003) 860-863.

[19] X.Y. Liu, H. Guo, Y. Liu, S. Ye, M.Y. Peng, Q.Y. Zhang, Thermal quenching and energy transfer in novel Bi3+/Mn2+ co-doped white-emitting borosilicate glasses for UV LEDs, Journal of Materials Chemistry C 4(13) (2016) 2506-2512.

[20] M. Takesue, H. Hayashi, R.L. Smith, Thermal and chemical methods for producing zinc silicate (willemite): A review, Progress in Crystal Growth and Characterization of Materials 55(3) (2009) 98-124.

[21] T. Gotoh, M. Jeem, L. Zhang, N. Okinaka, S. Watanabe, Synthesis of yellow persistent phosphor garnet by mixed fuel solution combustion synthesis and its characteristic, Journal of Physics and Chemistry of Solids 142 (2020) 109436.

[22] E. Carlos, R. Martins, E. Fortunato, R. Branquinho, Solution Combustion Synthesis: Towards a Sustainable Approach for Metal Oxides, Chemistry-a European Journal 26(42) (2020) 9099-9125.

[23] Z. Yin, S. Li, X. Li, W. Shi, W. Liu, Z. Gao, M. Tao, C. Ma, Y. Liu, A review on the synthesis of metal oxide nanomaterials by microwave induced solution combustion, RSC Advances 13(5) (2023) 3265-3277.

[24] Z. Yang, X. Li, Y. Yang, X. Li, The influence of different conditions on the luminescent properties of YAG:Ce phosphor formed by combustion, Journal of Luminescence 122-123 (2007) 707-709.

[25] F. Siddique, S. Gonzalez-Cortes, A. Mirzaei, T. Xiao, M.A. Rafiq, X. Zhang, Solution combustion synthesis: the relevant metrics for producing advanced and nanostructured photocatalysts, Nanoscale 14(33) (2022) 11806-11868.

[26] A.M. Pires, M.R. Davolos, Luminescence of Europium(III) and Manganese(II) in Barium and Zinc Orthosilicate, Chem. Mat. 13(1) (2001) 21-27.

[27] K.-H. Klaska, J.C. Eck, D. Pohl, New investigation of willemite, Acta Crystallographica Section B 34(11) (1978) 3324-3325.

[28] T. Hasegawa, Y. Nishiwaki, F. Fujishiro, S. Kamei, T. Ueda, Quantitative Determination of the Effective Mn4+ Concentration in a Li2TiO3:Mn4+ Phosphor and Its Effect on the Photoluminescence Efficiency of Deep Red Emission, ACS Omega 4(22) (2019) 19856-19862.

[29] E. Beyreuther, S. Grafström, L.M. Eng, C. Thiele, K. Dörr, XPS investigation of Mn valence in lanthanum manganite thin films under variation of oxygen content,

Physical Review B 73(15) (2006) 9.

[30] M.C. Biesinger, B.P. Payne, A.P. Grosvenor, L.W.M. Lau, A.R. Gerson, R.S. Smart, Resolving surface chemical states in XPS analysis of first row transition metals, oxides and hydroxides: Cr, Mn, Fe, Co and Ni, Appl. Surf. Sci. 257(7) (2011) 2717-2730.

[31] Y. Hao, Y.-H. Wang, Synthesis and photoluminescence of new phosphors M2(Mg, Zn)Si2O7:Mn2+ (M=Ca, Sr, Ba), Materials Research Bulletin 42(12) (2007) 2219-2223.

[32] C.E. Rivera-Enríquez, A. Fernández-Osorio, J. Chávez-Fernández, Luminescence properties of α- and β-Zn2SiO4:Mn nanoparticles prepared by a co-precipitation method, Journal of Alloys and Compounds 688 (2016) 775-782.

[33] Q. Lu, P. Wang, J. Li, Structure and luminescence properties of Mn-doped Zn2SiO4 prepared with extracted mesoporous silica, Materials Research Bulletin 46(6) (2011) 791-795.

[34] Y.Q. Jiang, J. Chen, Z.X. Xie, L.S. Zheng, Syntheses and optical properties of α- and β-Zn2SiO4: Mn nanoparticles by solvothermal method in ethylene glycol-water system, Mater. Chem. Phys. 120(2-3) (2010) 313-318.

[35] V. Boiko, J. Zeler, M. Markowska, Z. Dai, A. Gerus, P. Bolek, E. Zych, D. Hreniak, Persistent luminescence from Y3Al2Ga3O12 doped with Ce3+ and Cr3+ after X-ray and blue light irradiation, J. Rare Earths 37(11) (2019) 1200-1205.

[36] J. Ueda, M. Katayama, K. Asami, J. Xu, Y. Inada, S. Tanabe, Evidence of valence state change of Ce3+ and Cr3+ during UV charging process in Y3Al2Ga3O12 persistent phosphors, Opt. Mater. Express 7(7) (2017) 2471-2476.

[37] S.W.S. McKeever, ON THE ANALYSIS OF COMPLEX THERMO-LUMINESCENCE GLOW-CURVES-RESOLUTION INTO INDIVIDUAL PEAKS, Phys. Status Solidi A-Appl. Res. 62(1) (1980) 331-340.

[38] R. Chen, On the Calculation of Activation Energies and Frequency Factors from Glow Curves, Journal of Applied Physics 40(2) (1969) 570-585.

[39] R. Chen, Glow Curves with General Order Kinetics, Journal of The Electrochemical Society 116(9) (1969) 1254.

[40] V.E. Kafadar, T. Yeşilkaynak, R.E. Demirdogen, A.A. Othman, F.M. Emen, S. Erat, The effect of Dy3+ doping on the thermoluminescence properties of Ba2SiO4, 17(3) (2020) 1453-1459.

[41] G.F.J. Garlick, A.F. Gibson, The Electron Trap Mechanism of Luminescence in Sulphide and Silicate Phosphors, Proceedings of the Physical Society 60(6) (1948) 574.

[42] Z. Chen, W. Cui, K. Zhu, C. Zhang, C. Zuo, Y. Niu, Q. Wang, X. Yuan, G. Liu, Improved luminescence and afterglow emission from Mn2+/Si4+ co-doped AlN by combustion synthesis method, Journal of Alloys and Compounds 883 (2021) 160745.